\documentclass[sigconf]{acmart}

\usepackage{multirow}
\usepackage{subcaption}
\AtBeginDocument{%
  }

\copyrightyear{2026}
\acmYear{2026}
\setcopyright{cc}
\setcctype{by}
\acmConference[RecSys '26]{20th ACM Conference on Recommender Systems}{September 27-October 02, 2026}{Minneapolis, MN, USA}
\acmBooktitle{20th ACM Conference on Recommender Systems (RecSys '26), September 27-October 02, 2026, Minneapolis, MN, USA}
\acmDOI{10.1145/3773078.3831891}
\acmISBN{979-8-4007-2284-4/2026/09}

\begin{document}

\title{Advancing Relevance Measurement with Vision–Language Models for Web-Scale Search}


\author{Han Wang}
\authornote{Equal contribution}
\orcid{0009-0005-5134-2136}
\affiliation{%
  \institution{Pinterest}
  \city{San Francisco, CA}
  \country{USA}}
\email{hanwang@pinterest.com}

\author{Alex Whitworth}
\authornotemark[1]
\affiliation{%
  \institution{Pinterest}
  \city{San Francisco, CA}
  \country{USA}}
\email{awhitworth@pinterest.com}

\author{Pak Ming Cheung}
\affiliation{%
  \institution{Pinterest}
  \city{San Francisco, CA}
  \country{USA}}
\email{pcheung@pinterest.com}

\author{Zhenjie Zhang}
\affiliation{%
  \institution{Pinterest}
  \city{San Francisco, CA}
  \country{USA}}
\email{zhenjiezhang@pinterest.com}

\author{Krishna Kamath}
\authornote{Work done at Pinterest}
\affiliation{%
  \institution{Pinterest}
  \city{San Francisco, CA}
  \country{USA}}
\email{krishna.kamath@gmail.com}

\author{Xi Chen}
\affiliation{%
  \institution{Pinterest}
  \city{San Francisco, CA}
  \country{USA}}
\email{xichen@pinterest.com}

\author{Roberto Konow}
\affiliation{%
  \institution{Pinterest}
  \city{San Francisco, CA}
  \country{USA}}
\email{rkonow@pinterest.com}

\author{Kurchi Subhra Hazra}
\affiliation{%
  \institution{Pinterest}
  \city{San Francisco, CA}
  \country{USA}}
\email{ksubhrahazra@pinterest.com}

\renewcommand{\shortauthors}{Wang et al.}

\begin{abstract}
Relevance evaluation plays a crucial role in personalized search systems, serving as a guardrail alongside user engagement metrics to ensure that search results align with user queries and intent. While human annotation is the traditional method for relevance evaluation, its high cost and long turnaround time limit its scalability. 
In this work, we present a VLM-based automated relevance evaluation pipeline deployed within Pinterest Search for online A/B experiments. We rigorously validate the alignment between VLM-generated judgments and human annotations, demonstrating that VLMs can provide reliable relevance measurement for experiments while greatly improving the evaluation efficiency. Leveraging VLM-based labeling further unlocks opportunities to expand the query set, optimize sampling design, and efficiently assess a wider range of search experiences at scale. This approach leads to higher-quality relevance metrics and significantly reduces the Minimum Detectable Effects (MDEs) in online experiment measurements.
\end{abstract}




\maketitle

\section{Introduction}

Search relevance measures how well search results align with a user's query. For personalized search systems, evaluating relevance is crucial to ensure that displayed results are pertinent to the user's information needs, rather than over-relying on prior engagement. 
Online A/B experiments are widely used in industry to measure the impact of ranking changes. 
For search systems, it is essential to track changes in  
semantic relevance alongside engagement metrics. 
Engagement metrics can be collected from real-time user actions (e.g., click, save), but user engagement does not always reflect the semantic relevance of the search feed: position, presentation, and attention effects all influence user actions independently of content quality. A personalization change may therefore drive strong engagement gains while introducing irrelevant content into top slots, degrading overall search relevance. 
Relevance evaluation therefore serves as a guardrail to detect such tradeoffs. 

Relevance evaluation typically relies on human annotations. However, this approach suffers from high costs, long turnaround time, and limited scalability, which can ultimately compromise the quality of relevance metrics. In recent years, the emergence of Large Language Models (LLMs) has sparked growing interest in automating relevance assessment in the Information Retrieval (IR) community. There have been several works exploring the feasibility of using LLMs to supplement laborious human labeling efforts \cite{faggioli2023perspectives,thomas2024large,upadhyay2024umbrela}. 
Extending LLM-based evaluation to multimodal settings, recent advances in Vision-Language Models (VLMs) have enabled native, long-context multimodal reasoning \cite{team2024gemini,bai2025qwen3}, making them well-suited for automated relevance evaluation \cite{yang2024toward}.

At Pinterest, search is one of the key surfaces where users discover content that meets their information needs. 
Beyond user engagement metrics, we also track semantic relevance of top slots to ensure a high-quality search experience. In this paper, we present our methodology for leveraging VLMs to assess semantic relevance in online A/B experiments at Pinterest Search. To address the multimodal nature of Pinterest content, we fine-tune open-source VLMs on relevance prediction tasks using human-annotated labels, then apply the fine-tuned VLMs to label search results in online A/B experiments.\footnote{A preliminary, non-archival version of this work appeared as a workshop paper \cite{wang2025llm}, which used a text-only model. The present paper extends it with a VLM-based multimodal pipeline, revises and expands the reported results, and includes additional ablation studies.} 
This approach not only significantly reduces labeling costs and achieves over $20\times$ improvement in labeling turnaround time, but also improves metric quality by scaling up query sets and refining sampling design. 
Since launch, the system has scaled to more than $4\times$ the relevance measurement jobs, 
enabling broader and more frequent relevance evaluation across Pinterest Search.
The primary contribution of this work lies in the end-to-end design and deployment of a VLM-based relevance evaluation pipeline for live A/B experimentation, rather than in a novel modeling architecture. To the best of our knowledge, this is the first work to comprehensively address this problem in an industry search system. We hope this work offers practical insights for industry practitioners seeking to improve the efficiency of relevance measurement. 

Our contributions are as follows:
\begin{itemize}
    \item We present an end-to-end pipeline for VLM-based relevance evaluation in a production search system, 
    covering practical considerations and validation steps before deploying automated relevance labeling at scale.
    \item We highlight that VLM-based relevance assessment enables expanding the query set and improving the query sampling design, which leads to higher-quality relevance metrics and a $6\times$ reduction in Minimum Detectable Effect (MDE), substantially improving sensitivity for online experiment evaluation.
    \item We demonstrate that fine-tuned VLMs 
    produce high-quality relevance metrics closely aligned with human judgments. The mean error in the query-level $sDCG@K$ metric (defined in Section \ref{seq:rel_measure_with_vlm}) remains within 0.03, and paired differences exhibit even lower bias, confirming robustness for A/B experiment evaluation.
\end{itemize}

\section{Related Work}

Recent studies have shown the potential of using LLMs for automating relevance assessments in Information Retrieval (IR), as part of the broader “LLM-as-a-judge” paradigm \cite{gu2024survey,zheng2023judging}. \citet{thomas2024large} and \citet{upadhyay2024umbrela} demonstrated that LLMs can accurately predict searcher preferences and generate human-quality relevance labels. A significant line of work investigates prompting techniques for LLM-based relevance labeling, including zero-shot \cite{upadhyay2024umbrela,arabzadeh2025benchmarking,upadhyay2024large} and few-shot \cite{macavaney2023one,pires2025expanding} approaches. 
Beyond text-only settings, \citet{yang2024toward} benchmarked VLMs for zero-shot relevance evaluation on image-text retrieval tasks. More recently, \citet{meng2025query} and \citet{abbasiantaeb2024can} showed that fine-tuning open-source LLMs on human-labeled relevance judgments yields more reliable predictions than few-shot prompting with significantly larger models. 

Unlike prior work that primarily compares LLMs/VLMs and prompting strategies \cite{meng2025query,abbasiantaeb2024can} or explores human-LLM collaboration \cite{shankar2024validates},
our work presents a real-world deployment of VLM-based relevance measurement within an industry-scale A/B experimentation pipeline at Pinterest Search. While VLMs have been applied to enhance multimodal representations for retrieval and ranking \cite{giahi2025vl,ye2023query}, our work addresses the complementary problem of automated relevance evaluation. 

\section{Methodology}

\subsection{Problem Statement}\label{sec:problem_statement}

At Pinterest, we measure the semantic relevance between search queries and Pins\footnote{Pins on Pinterest are rich multimedia entities that feature images, videos, and other content, often linked to external webpages or blogs.} using a 5-level rating scale: Highly Relevant (L5), Relevant (L4), Marginally Relevant (L3), Irrelevant (L2), and Highly Irrelevant (L1). In online experiments, assessing relevance across experimental groups is essential for detecting the impact of ranking changes on top-slot relevance, beyond what engagement metrics alone can reveal. 

The relevance measurement has historically been limited by the low availability of human labels and the high per-label cost. This led to measurement designs and sample sizes that could only detect large topline metric movements, but were insufficient to measure heterogeneous treatment effects or small topline effects. Next, we describe how we leverage VLMs to scale relevance labeling and address these measurement bottlenecks.

\subsection{Fine-tuned VLM as Relevance Model}

We fine-tune the open-source Qwen3-VL models \cite{bai2025qwen3} on human-annotated query-Pin pairs to optimize relevance prediction performance. 
The training dataset contains approximately 0.8 million human-annotated pairs, with model selection performed on a separate held-out set of approximately 20K pairs.
To support search queries across multiple languages, we leverage the cross-lingual transfer capabilities of multilingual Qwen3-VL. 
We prompt the model to generate a relevance score from "1" to "5" following the 5-level rating guideline (Section \ref{sec:problem_statement}), minimizing cross-entropy loss during training. At inference, we take the argmax over the logits of output tokens "1" through "5" as the predicted relevance level.

\begin{figure}[h!]
  \centering
  \includegraphics[width=0.8\linewidth]{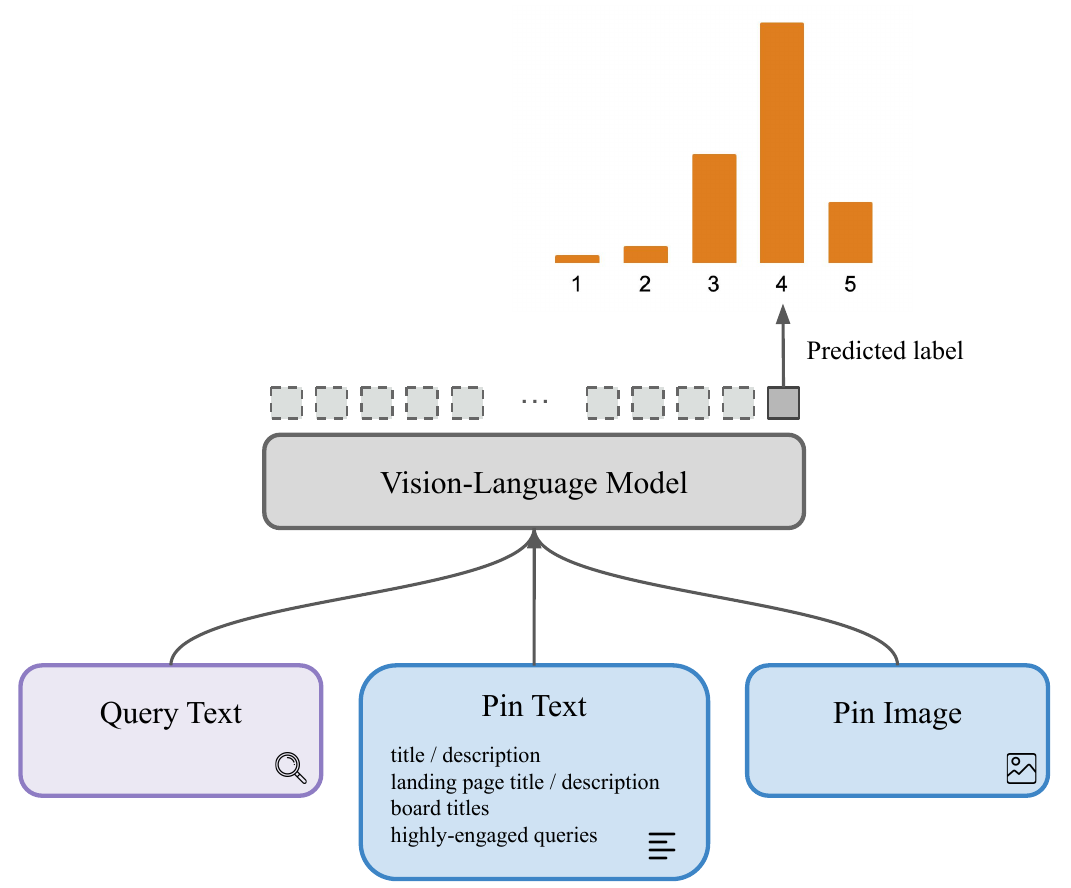}
    \caption{The fine-tuned VLM takes query text, Pin text metadata, and Pin image as input, and outputs a graded relevance label on a 5-point scale via argmax over the output logits.}
  \Description{Architecture of the fine-tuned VLM relevance teacher. The model takes three inputs: query text, Pin text metadata (title, description, landing page title and description, board titles, and engagement-derived queries), and Pin image. The VLM predicted label is the argmax over the logits of output tokens "1" through "5".}
  \label{fig:model_architecture}
\end{figure}

Beyond the Pin image, we leverage a comprehensive set of textual metadata to represent each Pin: title and description, landing page title and description, titles of user-curated boards the Pin has been saved to, and 
historically highly-engaged queries for the Pin. 
Although derived from engagement signals, these queries are aggregated across all users and thus serve as complementary content annotations.
Together, these textual features and the Pin image form a rich multimodal representation for relevance prediction, as illustrated in Figure \ref{fig:model_architecture}. 

\subsection{Stratified Sampling Design}

VLM labeling significantly reduces both the cost and turnaround time of relevance labeling, enabling much larger sampling designs. 
We therefore propose a stratified query sampling design that enables measurement of heterogeneous treatment effects and reduces MDEs by 6 times. Prior to VLM labeling, stratified query sampling with human annotations was impractical, as representing each stratum adequately required a large number of queries. 

Stratification plays an important role in sampling-based measurement. First, stratification ensures the sample population is representative of the whole population. In addition, when the strata are chosen such that 
stratum means are heterogeneous, 
variance reduction can be achieved \cite{miratrix2013adjusting}. Formally, let the relevance score be a random variable $Y$ with mean $\mu$ and variance $\sigma^2$. The variance of the sample mean $\bar Y$ is $V(\bar Y) = \sigma^2 / N$, where $N$ is the total sample size. 
When $K$ strata are defined such that each stratum has a different mean, $V(\bar{Y})$ can be decomposed into within-strata and between-strata variance (see Equation \ref{eq:strata_variance}). Stratified sampling eliminates the between-strata term \cite{xie2016improving}, yielding: 
\begin{equation}
    V(\bar Y_{SRS})= \underbrace{\sum_{k=1}^K \frac{n_k}{N} \sigma_k^2}_{\text{within-strata}} + \underbrace{\sum_{k=1}^K \frac{n_k}{N} (\mu_k - \mu)^2}_{\text{between-strata}} \geq \sum_{k=1}^K \frac{n_k}{N} \sigma_k^2 = V(\bar Y_{strat}),
    \label{eq:strata_variance}
\end{equation}
where $n_k$, $\mu_k$, $\sigma_k^2$ denote the sample size, mean, and variance of each stratum. The resulting variance reduction directly translates to reduced MDEs. 

The primary source of variance in our relevance measurement system is between-query variation: strata defined on query characteristics show large mean differences driven by query intent, content quality, and inventory depth. Stratification directly targets this variance structure, yielding design effects (DEFF) well below 1.
To determine the query strata, we evaluated multiple stratification choices. While BERTopic \cite{grootendorst2022bertopic} achieved the largest variance reduction, we adopted an in-house query-to-interest model based on DistilBERT \cite{sanh2019distilbert} for its ease of integration with existing systems and scalability.

For completeness, we note why CUPED \citep{Deng2013ImprovingTS} is less suitable here. CUPED requires pre-treatment covariate observations, which in this context would mean restricting the sampling frame to query-user pairs with recent repeated queries, a substantial exclusion that would compromise the generalizability of resulting experimental estimates. Even where such observations exist, their utility is limited: Pinterest is a highly dynamic system in which engagement signals and inventory evolve continuously, making pre-treatment scores a weak proxy for current relevance. Finally, labeling repeated queries would double annotation cost. Stratified sampling avoids all three issues while targeting the dominant variance structure directly. 

\begin{figure*}[h!]
  \centering
  \includegraphics[width=0.9\linewidth]{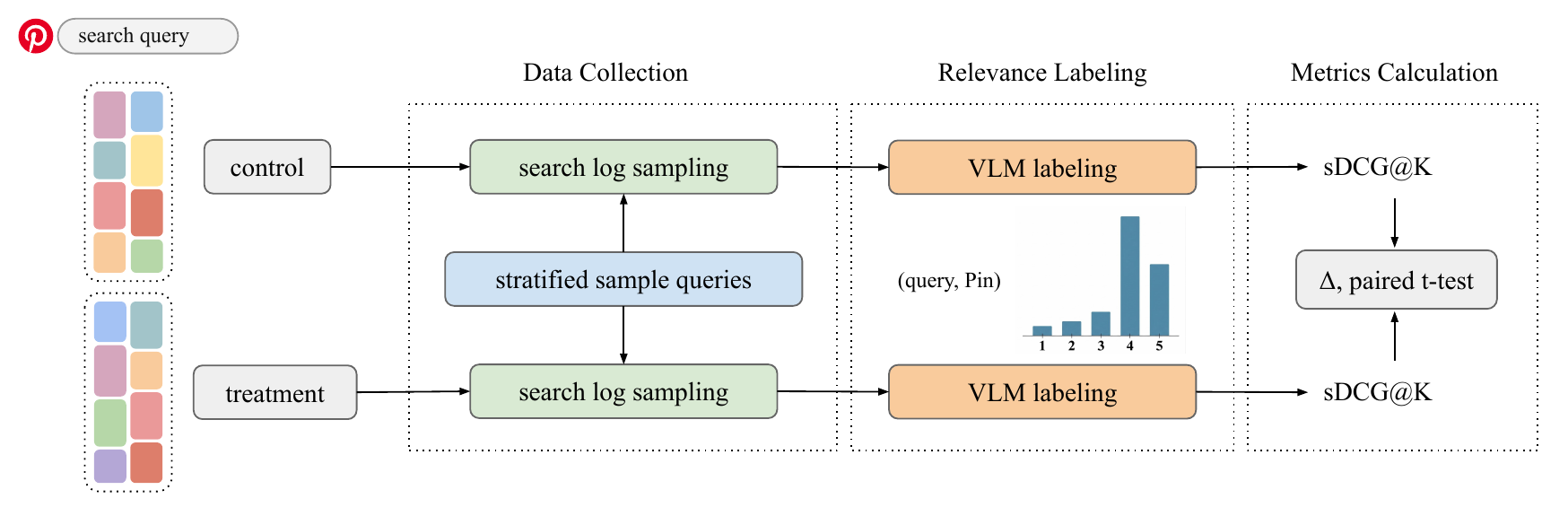}
  \caption{The VLM-based relevance measurement pipeline at Pinterest Search, consisting of search result sampling with stratified queries, VLM-based relevance labeling, and metric calculation for A/B experiment evaluation.}
  \Description{End-to-end pipeline for VLM-based offline A/B evaluation. Given a search query, search logs from the control and treatment arms are sampled using a shared set of stratified sample queries. Each sampled (query, Pin) pair is labeled by the VLM to produce a 5-level relevance score, from which per-query sDCG@K is computed for both arms. The metric delta between arms is then assessed via a paired t-test to determine statistical significance.}
  \label{fig:relevance_measurement}
\end{figure*}

\subsection{Relevance Measurement with VLMs}
\label{seq:rel_measure_with_vlm}

To measure the relevance impact of an A/B experiment on search ranking, we take a stratified sample of paired search queries from control and treatment experiment groups\footnote{Excluding queries that appear in only one experimental group is a deliberate restriction of the target population, not a structural bias toward high-volume queries.}, ensuring that the sample is representative of the overall query distribution \cite{thompson2012sampling}. 
The use of paired samples blocks between-query differences, an important source of variation in experiment measurement.

For each query in our paired sample, we retain the top $K$ search results and generate VLM-based relevance labels. We then compute $sDCG@K$ for each query and aggregate query-level metrics to derive experiment-level topline metrics. The $sDCG@K$ metric is a variant of the standard $nDCG@K$, where we assume an infinite supply of highly relevant (L5) documents (see Equation \ref{eq:sdcg}). We use $K=25$ throughout our evaluation. 
\begin{equation}
\label{eq:sdcg}
    sDCG@K=\frac{\sum_{k=1}^K l_k / \log_2(1+k)}{\sum_{k=1}^K 5 / \log_2(1+k)}, \text{ }l_k\in\{1,2,3,4,5\}.
\end{equation}

Finally, we calculate heterogeneous treatment effects stratified by query popularity and query interest category (e.g., beauty, women’s fashion, art), applying a Benjamini-Hochberg procedure \cite{benjamini1995controlling} to control the false discovery rate. 
The full VLM-based relevance measurement procedure is illustrated in Figure \ref{fig:relevance_measurement}. 

\section{Results}
\label{section:results}

We fine-tune the open-source Qwen3-VL-4B model for automated relevance evaluation. The fine-tuned model labels hundreds of thousands of query-Pin pairs within 
two hours
on a single A100 GPU, 
achieving over $20\times$ improvement in labeling turnaround time compared to human annotation. 

We evaluate the fine-tuned model across three research questions (RQs): 
\begin{itemize}
    \item \textbf{RQ1}: Can VLM-based relevance assessment provide metrics that reliably align with human judgments for online A/B experiment measurements?
    \item \textbf{RQ2}: Does VLM-based relevance assessment offer greater metric sensitivity compared to human labeling?
    \item \textbf{RQ3}: Is multilingual VLM-based relevance assessment effective for non-English queries? 
\end{itemize}

\subsection{RQ1: Alignment with Human Labels}
\label{sec:rq1_us}

We validate the metrics derived from VLM labeling by comparing against human labels using query-Pin pairs from live experiment traffic. The human labels were collected by instructing annotators to rate the relevance of each query-Pin pair using the same 5-level rating scale. 
The exact match rate between VLM and human labels is 82.9\%, with 94.2\% of ratings differing by no more than one point, which is comparable to the inter-annotator agreement. Quadratic Weighted Kappa (QWK) \citep{Cohen1968WeightedKN} is a chance-corrected ordinal agreement statistic that penalizes larger disagreements more heavily. Our model achieves a QWK of 0.507, indicating good alignment between VLM-generated and human relevance labels. 

Following established practices in the literature \cite{meng2025query,abbasiantaeb2024can}, we calculate the rank-based correlation metrics, Kendall’s $\tau$ and Spearman’s $\rho$, to assess agreement between VLM and human rankings on the query-level sDCG@K metric. Overall, we achieve Kendall’s $\tau=0.534$ and Spearman’s $\rho=0.668$, indicating strong alignment between VLM and human rankings across all popularity segments. 

\begin{table}[h!]
  \caption{Alignment between VLM and human labels in query-level sDCG@K for relevance evaluation across different query popularity segments in the US market. }
  \label{tab:us_label_alignment}
  \scalebox{1.0}{\begin{tabular}{l|ccc|ccc}
    \toprule
    \multirow{2}{4em}{Segment} &  \multicolumn{3}{c|}{Query-level Error}  & \multicolumn{3}{c}{\textbf{Paired Difference Error}} \\
    &  Mean & P10 & P90 & {Mean} & {P10} & {P90} \\
    \midrule
     Overall  & 0.021	 &  -0.011 & 0.073 &  0.000 & -0.013 & 0.013 \\
    \midrule
    Head   & 0.018	 & 0.000 &  0.045 & 0.001 & -0.012 & 0.009 \\
    Torso   & 0.022	 & 0.000  & 0.060 & 0.000 & -0.009 & 0.008 \\
    Tail  & 0.020 & -0.003  & 0.015 & 0.000 & -0.009& 0.010\\	 
    Single  & 0.021 & -0.042  & 0.084 & 0.001  &-0.016 & 0.017 \\
  \bottomrule
\end{tabular}}
\end{table}

\begin{figure}[h]
  \centering
  \includegraphics[width=0.9\linewidth]{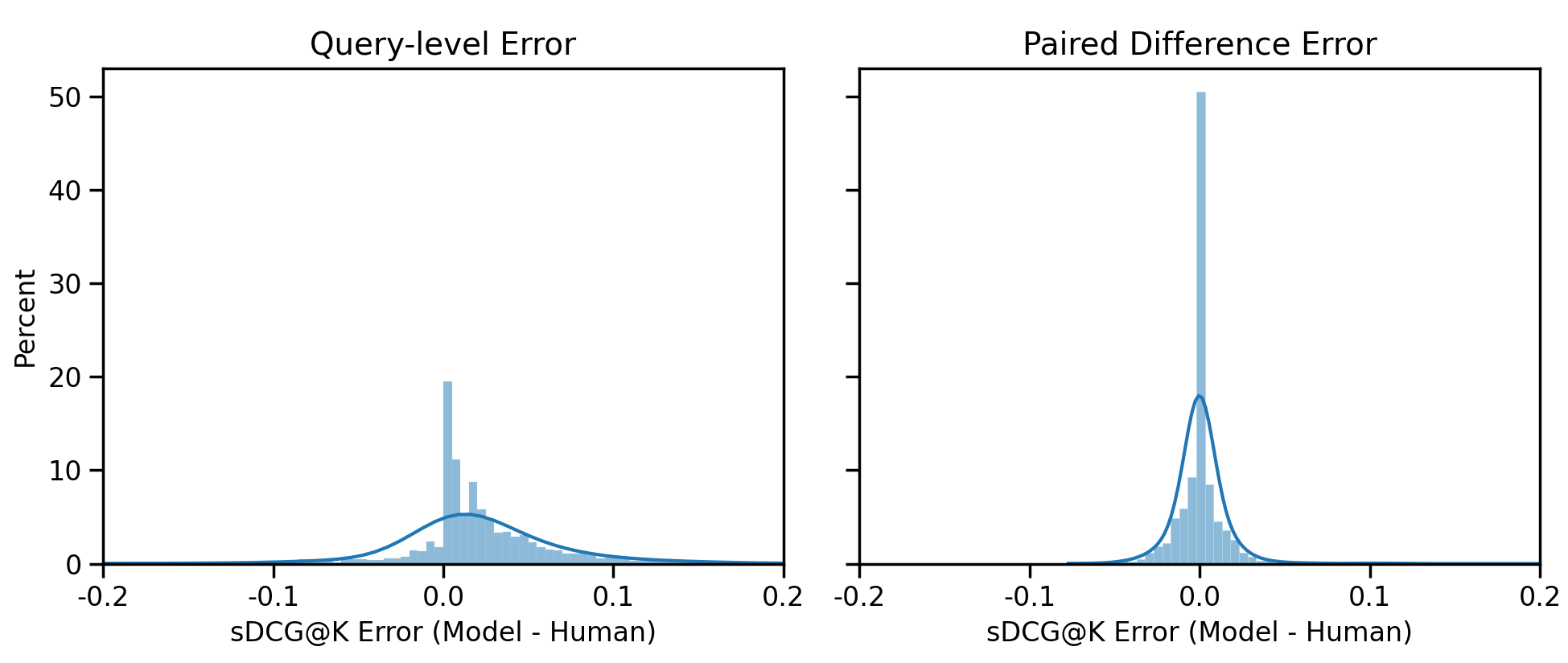}
  \caption{Query-level $\boldsymbol{sDCG@K}$ error distribution for single group (left) and paired differences (right) in the US market relevance evaluation. }
  \Description{Distribution of sDCG@K errors between VLM-derived and human-derived labels. (Left) Query-level error. (Right) Paired difference error. Both distributions are tightly centered around zero; the paired difference error exhibits a substantially narrower spread, indicating that systematic label biases cancel out in the paired A/B comparison and confirming the robustness of the evaluation design.}
  \label{fig:us_error}
\end{figure}

Beyond ranking alignment, we further validate the query-level sDCG@K error distribution, defined as VLM-derived sDCG@K minus human-derived sDCG@K. We focus on query-level rather than experiment-level error because each experiment is a single 
estimate with limited statistical power, whereas query-level evaluation substantially increases sample size and operates at the actual estimation unit. 
To assess the performance across query popularity levels, we categorize the queries into four popularity segments based on search volume \cite{jain2010organizing}: head, torso, tail, and single\footnote{Single queries are defined as those that receive fewer than 10 searches.}. 
As shown in Table \ref{tab:us_label_alignment}, the mean error is below 0.03 across all popularity segments, with P10 and P90 within [-0.1, 0.1], indicating low bias and well-bounded variance.  
Since A/B evaluation relies on paired metric differences, we additionally validate paired difference errors. 
According to Table \ref{tab:us_label_alignment}, the mean paired difference error has negligible magnitude across all segments, and notably eliminates the slight positive bias observed in query-level errors, confirming that the paired experimental design is robust to VLM label bias. 
We visualize both error distributions in Figure \ref{fig:us_error}. 
Both are tightly centered around 0, with paired differences exhibiting a tighter distribution, supporting the use of VLM-based labeling for reliable online A/B experiment evaluation.

\begin{table*}[!h]
  \caption{Performance of VLMs and text-only models on automated relevance evaluation in the US market. Relative gains over the XLM-RoBERTa-large baseline are shown for Accuracy, QWK and rank correlation metrics. }
  \label{tab:model_compare}
  \scalebox{0.9}{\begin{tabular}{l|cc|cc|ccc|ccc}
    \toprule
    \multirow{2}{4em}{Model} 
    & \multirow{2}{4em}{Accuracy} 
    & \multirow{2}{3em}{QWK}
    & \multicolumn{2}{c|}{sDCG Rank Correlation} & \multicolumn{3}{c|}{Query-level Error} & \multicolumn{3}{c}{{Paired Difference Error}}  \\
    &  & & Kendall's $\tau$ & Spearman's $\rho$ & Mean & P10 & P90 & Mean & P10 & P90 \\
    \midrule
    XLM-RoBERTa-large (text-only) & 0.808 
    & 0.429
    & 0.490 
    & 0.626 & 0.015 & -0.032 & 0.074 & 0.000 & -0.015 & 0.015 \\
    \midrule
    Qwen3-VL-2B 
    &0.824 (+2.0\%)  
    & 0.490 (+14.2\%)
    & 0.522 (+6.5\%) & 0.658 (+5.1\%) & 0.019 &  -0.016& 0.074 & 0.000 & -0.013 & 0.014 \\
    Qwen3-VL-4B 
    & \textbf{0.829 (+2.6\%)} 
    & 0.507 (+18.2\%)
    & \textbf{0.534 (+9.0\%)} & 0.668 (+6.7\%) & 0.021 & -0.011 & 0.073 & 0.000 & -0.013 & 0.013\\	
    Qwen3-VL-8B 
    & \textbf{0.829 (+2.6\%)} 
    & \textbf{0.514 (+19.8\%)}
    & {0.532 (+8.6\%)} & \textbf{0.670 (+7.0\%)} & 0.018 & -0.016 & 0.069 & 0.000 & -0.013 & 0.013 \\	
  \bottomrule
\end{tabular}}
\end{table*}

\begin{table*}[h!]
  \caption{Alignment between VLM and human labels in query-level sDCG@K for relevance evaluation across France (FR), Germany (DE), and Brazil (BR) markets. }
  \label{tab:fr_de_label_alignment}
  \scalebox{1.0}{\begin{tabular}{l|cc|ccc|ccc}
    \toprule
    \multirow{2}{4em}{Country} & \multicolumn{2}{c|}{sDCG Rank Correlation} & \multicolumn{3}{c|}{Query-level Error} & \multicolumn{3}{c}{{Paired Difference Error}}  \\
    & Kendall's $\tau$ & Spearman's $\rho$ & Mean & P10 & P90 & Mean & P10 & P90 \\
    \midrule
    FR & 0.430 & 0.544 & 0.035 & -0.002 & 0.098 & -0.002 & -0.051& 0.046\\
    DE & 0.362 & 0.460 & 0.023 & -0.021 & 0.087 & -0.004&-0.049 & 0.044\\
    BR & 0.409 & 0.493 & 0.040 & 0.000 & 0.140 & 0.000 & -0.065 & 0.070 \\
  \bottomrule
\end{tabular}}
\end{table*}

We next evaluate Qwen3-VL across different model sizes and compare against a text-only XLM-RoBERTa-large baseline augmented with BLIP-generated image captions \cite{wang2024improving}, to assess the accuracy-efficiency tradeoff and the value of visual inputs.
Results for the US market are summarized in Table \ref{tab:model_compare}. 
Qwen3-VL-4B and Qwen3-VL-8B achieve comparable performance across all metrics, with near-identical accuracy and rank correlations within 0.002 of each other.  
Notably, the text-only XLM-RoBERTa model has a low mean query-level error, but its wider P10–P90 range indicates higher variance, meaning more queries where model and human judgments diverge substantially. 
For reliable relevance measurement, a tight error distribution is preferable. 
We therefore adopt Qwen3-VL-4B as our production model, as it offers strong relevance prediction quality, a tight error distribution, and lower computational cost than the 8B variant.

\subsection{RQ2: Metrics Sensitivity and MDEs}

We evaluate the impact of these changes on experiment sensitivity by measuring MDEs for our experimentation system. 
The MDE is the smallest change in a metric that an experiment can reliably detect given the sample size, statistical power, and significance level chosen for the test. 
Since most online experiments produce small effect sizes, achieving small MDEs is essential for detecting meaningful improvements and maintaining experimentation velocity.

\begin{table}[h!]
  \caption{Improvement in metric sensitivity (MDE) with proposed sampling design and estimator.}
  \label{tab:mde}
  \scalebox{0.95}{
  \begin{tabular}{ccccc}
    \toprule
    Sample Type & Estimator & $n$ & $\hat \sigma$ & Reduction in $\hat \sigma$ \\
    \midrule
    SRS & SRS & 2000 & 0.184 & - \\
    Stratified & Stratified & 2000 & 0.096 & 52\% \\
    Stratified & SRS & 5000 & 0.094 & 51\% \\
    Stratified & Stratified & 5000 & 0.061 & 67\% \\
  \bottomrule
\end{tabular}}
\end{table}

Before the introduction of VLM labeling, relevance measurement had large MDEs, typically ranging from 1.3\% to 1.5\%. These large MDEs were primarily the result of the constraints on our sampling designs imposed by the high cost and turnaround time of human labeling. 
The introduction of VLM labeling removed these constraints, enabling us to redesign our sampling approach. We increased our sample sizes, moved from simple random sampling (SRS) to stratified sampling, and adopted a stratified sampling estimator \cite{thompson2012sampling}. Optimal allocation \cite{neyman_allocation} is used to allocate sample units to strata, which are defined as the cross-product of query interest category and popularity segment. These changes enabled us to reduce our MDEs to $\leq 0.25\%$.

To attribute the MDE reduction to each source of change, we follow a standard derivation of MDE \cite{ds_power} as a function of the number of queries in the sample ($n$), metric mean ($\hat \mu$), and variance ($\hat \sigma^2$), assuming standard values for $\alpha = 0.05$ and $\beta = 0.8$,
\begin{equation}
    \mathrm{MDE} = \mathrm{Lift}\% \mathrm{Detectable}=
    \frac{(z_{1-\alpha/2} + z_{\beta}) \times \sqrt{2\hat{\sigma}^2/ n}}{\hat{\mu}}.
    \label{mde_eqn}
\end{equation}
We can therefore decompose MDE reduction into contributions from variance reduction and sample size increase. We present these results in Table \ref{tab:mde}. The largest share of MDE reduction is attributable to variance reduction from stratification, consistent with our observation at Pinterest that most variance in relevance occurs across queries,
largely due to differences in query interest and popularity.

Beyond metric quality, VLM labeling also achieves over a 20-fold improvement in turnaround time and a greater than 99\% reduction in per-label cost, as shown in Table \ref{tab:label_cost_compare}.

\begin{table}[h!]
  \caption{Comparison of label volume, per-label cost, and turnaround time before and after adopting VLM-based labeling.}
  \label{tab:label_cost_compare}
  \scalebox{0.88}{
  \begin{tabular}{cccc}
    \toprule
     & Label Volume & Per-label Cost (\$) & Turnaround Time \\
    \midrule
    Baseline & 
    $O(10K)$
    & 0.1 & 2 days \\
    VLM Labeling & 
    $O(100K)$
    & $2\times 10^{-5}$ (-99.98\%) & 2 hours (-96\%)  \\
  \bottomrule
\end{tabular}}
\end{table}

\subsection{RQ3: Performance on Non-English Queries}

Since the fine-tuning data consists predominantly of English query-Pin pairs, careful validation is required before extending VLM-based assessment to non-English queries. For this analysis, we focus on the France (FR), Germany (DE), and Brazil (BR) markets as representative non-English markets, using query-Pin pairs from live experiment traffic for validation.

The query-level metric alignment is summarized in Table \ref{tab:fr_de_label_alignment}. 
While Kendall’s $\tau$ and Spearman’s $\rho$ are lower than those observed for English queries (Table \ref{tab:us_label_alignment}), they are generally interpreted as moderate-to-strong correlations according to existing literature \cite{yang2024toward,meng2025query}.  
Notably, query-level errors show a slight positive bias; however, this bias is effectively eliminated in paired differences, with errors tightly concentrated around 0. 
These results suggest that the VLM-based relevance assessment generalizes effectively to non-English markets.

\section{Conclusion and Future Work}

In this work, we present a VLM-based relevance labeling framework for generating query-level relevance metrics in online A/B experiment evaluation. We demonstrate that fine-tuned VLMs achieve low bias on query-level $sDCG@K$ metrics and paired differences, with strong alignment to human annotations across query popularity segments. Transitioning to VLM-based relevance assessment enables us to scale up the evaluation query set and redesign the sampling strategy, yielding a $6\times$ reduction in MDEs and significantly improved detection of relevance shifts. We have successfully deployed this approach at Pinterest Search, significantly reducing manual annotation costs and achieving over $20\times$ improvement in labeling turnaround time while maintaining metric reliability. With the efficiency boost, the relevance measurement job volume has grown more than $4\times$ since launch, reflecting both the efficiency of the VLM-based pipeline and the growing demand for automated relevance measurement. 
Future work will extend VLM-based relevance measurement to broader multimodal search settings beyond text-to-image search, and improve multilingual capabilities to close the observed performance gap on non-English queries.

\begin{acks}
We would like to thank Kofi Boakye, Sheng-Min Shih, Bonnie Liu, Manas Pathak for their collaboration in developing the VLM-based relevance model; 
and Miguel Madera, Pedro Sanchez, Jorge Amigon, Francisco Navarrete for their contributions to the automated labeling integration.
\end{acks}

\bibliographystyle{ACM-Reference-Format}
\bibliography{main-ref}

\appendix

\end{document}